\documentclass[pra, 12 pt]{revtex4}
\usepackage[english]{babel}
\usepackage{amsmath}
\usepackage{epsfig}

\begin{document}

\title{Universality of phonon-roton spectrum in liquids and superfluidity of He II}

\author{V.B.Bobrov $^{1,2}$, S.A.Trigger$^{1,3}$, D.I. Litinski $^3$}

\address{$^1$ Joint\, Institute\, for\, High\, Temperatures, Russian\, Academy\,
of\, Sciences, 13/19, Izhorskaia Str., Moscow\, 125412, Russia;\\
$^2$ National Research University "MPEI"\,,
Krasnokazarmennaya str. 14, Moscow, 111250, Russia;\\
$^3$ Eindhoven  University of Technology, P.O. Box 513, MB 5600
Eindhoven, The Netherlands;\\
emails:\, vic5907@mail.ru,\;satron@mail.ru}

\begin{abstract}

Based on numerous experimental data on inelastic neutron and X-ray scattering in liquids, we assert that the phonon-roton spectrum of elementary excitations, predicted by Landau for superfluid helium, is a universal property of the liquid state. We show that the existence of the roton minimum in the spectrum of collective excitations is caused by the short-range order in liquids. Using the superfluidity criterion, we assume that one more branch of collective excitations should exist in He II, whose energy spectrum differs from the phonon-roton spectrum. \\

PACS number(s): 05.30.Jp, 03.75.Kk, 03.75.Nt\\

\end{abstract}

\maketitle

\section{Introduction}

In phenomenological theory of superfluidity [1], Landau first put into consideration the notion of quasiparticles as quantized collective excitations. Based on an analysis of available experimental data on the specific heat and second sound, Landau predicted the so-called phonon-roton energy spectrum of such excitations $E_{p-r}(p)$ in superfluid helium, and formulated the criterion of superfluidity disappearance above a certain critical flow velocity [1,2]. To test the Landau hypothesis, Cohen and Feynman [3] proposed to determine the collective excitation spectrum by the maxima $E_{max}(p) $ in the dynamic structure factor $S(p, E)$ of liquid. The values of the function $S(p, E) $ can be experimentally determined by the cross section of inelastic neutron scattering (INS) in liquid, depending on the momentum $p $ and energy $E $. These ideas were implemented in quite a number of experimental works (see, e.g., [4,5]), results of which excellently confirmed the Landau's assumption on the shape of the spectrum of collective excitations in superfluid helium (see Fig. 1).

\begin{figure}[h]
\centering\includegraphics[width=6cm]{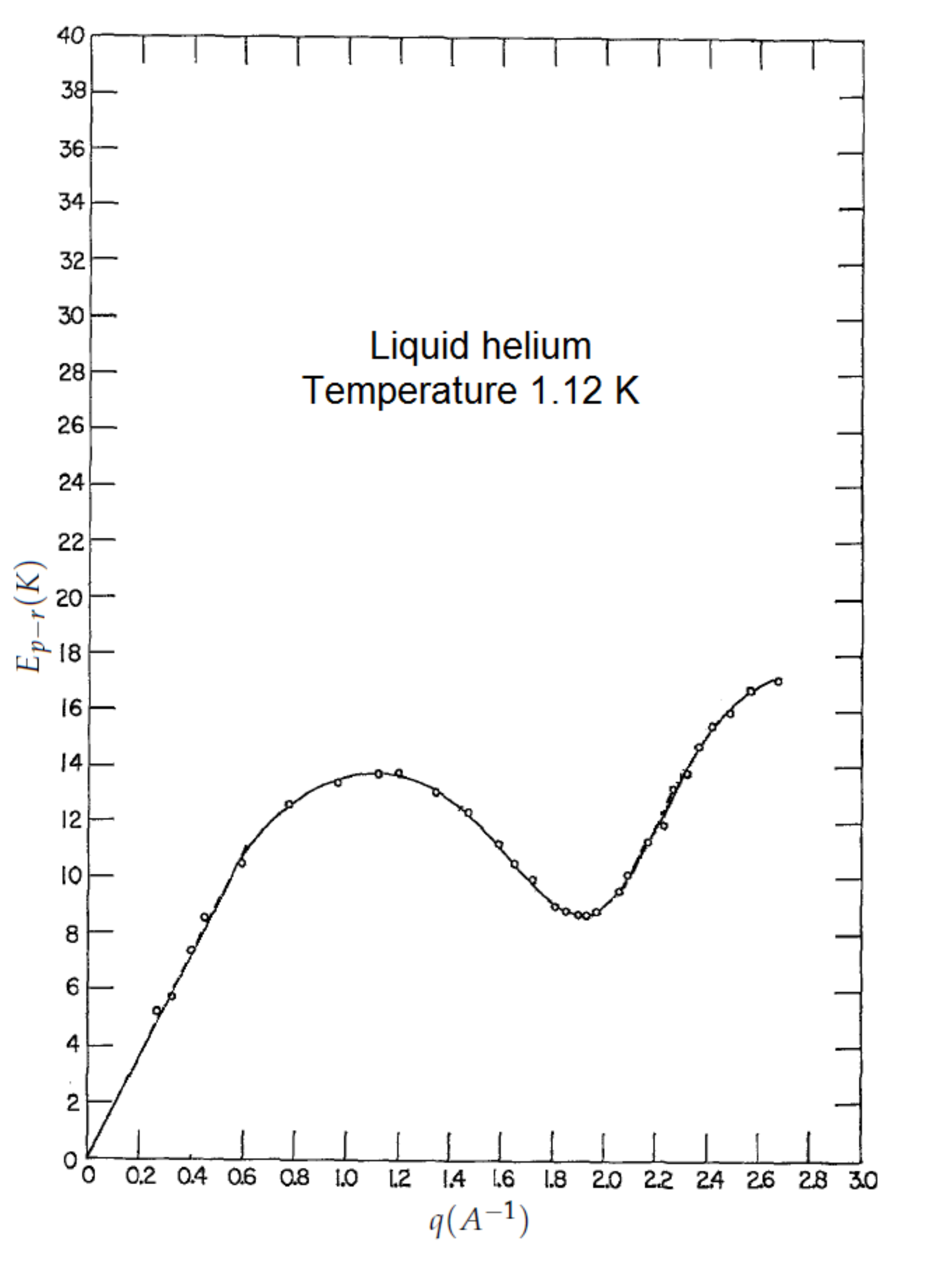}
\caption{{\protect\footnotesize {Confirmation of the predicted Landau [2] shape of excitations in superfluid helium: typical phonon-rhoton curve in helium at $T=1.12 K$ [5]. }}}\label{Fig.1}
\end{figure}

The capability of studying the collective dynamics in liquids using INS experiments was later complemented by experiments on inelastic X-ray scattering (IXS) [6,7].

We have stress that the experimental data explicitly show the existence of the peak $S(p,E)$ and the respective curve $E_{p-r}(p)$ of maxima position for various liquids (as we demonstrate below). At the same time, the Landau assumption about existence of rotons, formulated for superfluid HeII in [1,2] till now accepts very often as the specific and typical property of the superfluid state only (see, e.g., [8]).

The $E_{p-r}(p)$ curve is similar to phonon-roton excitations curve, predicted by Landau in superfluid helium. However, further interpretation of these maxima as well defined excitations, or quasiparticles, based on the specific theoretical models and assumptions about analytical shape of the function $S(p,E)$. The existence of quasiparticles means smallness of the respective damping in comparison with the characteristic frequency. At the same time maximum $S(p,E)$ can exist in the case of high damping also. Therefore, it is necessary to distinguish the curve of maxima in $S(p,E)$ and the statement about the existence of the well defined quasiparticles in liquids.
Moreover, the rigorous theoretical description of the collective excitations form in liquid, as in a system with strong interparticle interaction, is impossible in principle. Therefore, the existence of quasiparticles and the form of their spectrum is determined, in fact, by the chosen model for the liquid correlation functions. This choice leads to different interpretation of the experimental data.

The general concept of the Bose condensed systems (Onsager, Penrose, Ginzburg, Landau) based on finite limit of the density matrix $ \rho (\mid {\bf r-r'} \mid)_{\mid {\bf r-r'} \mid \rightarrow\infty} \rightarrow n_0$ cannot provide the description of the Bose systems in detail. In contrast with this concept, the Landau phenomenological theory does not use the notion about Bose condensation at all. In this sense the Landau phenomenological thermodynamic theory of liquid HeII based on the phonon-roton excitations can be considered as unexpected success. It is necessary to stress that Landau theory is valid only essentially below of transition temperature $T=T_\lambda$. The apparent achievement of the theory [2] is the similarity of the spectrum of quasiparticles suggested by Landau and the curve of the maxima of the dynamical structure factor, measured later on by the neutron and X-ray scattering.

Nevertheless, we will use below the term phonon-roton branch of excitations, taking into account that for the region of wave vectors where roton maximum exists the notion of quasiparticles is questionable.

The construction of the paper is following. In section II we collected the experimental data to show universality of the phonon-roton excitations in liquids. It is especially actually in the connection of the recent attempts to qualify these excitations as the specific excitations for superfluid HeII, which disappear above $T>T_\lambda$ [8]. In section III the theoretical background for the universality of the phonon-roton excitations in various liquids is formulated. In section IV we discuss the opportunity for existence of a new branch of excitations (helons), specific only for superfluid state and disappearing at $T=T_\lambda$ and establish similarities and distinguishes of helons with the rotons, assumed Landau in [1]. In section V (Conclusions) we discuss possible origin of the helons as the poles of the single-particle Green function, which can be different from the collective excitations in the retarded density-density correlation function.

\section{Universality of phonon-roton excitations: the selected experimental data}

To date, there are experimental data on maxima existence in a wide temperature range for many liquids with various physical properties. Below we present the collected experimental data to argue this statement. This argumentation seems necessary due to the wide spread opinion, mentioned above, that rotons exist only in superfluid state. The experimental data convincingly show that the phonon-roton spectrum is characteristic of not only superfluid helium, but is also a universal characteristic of the liquid state:

(A) Conductive liquids - liquid metals.
   Experimental data for various liquid metals are given in the reviews [9,10]. The analogy between the spectra of collective excitations in HeII and liquid metals was indicated in [11]. As an illustration, we present the results of later experiments with liquid titanium [12] (Fig.2).
Furthermore, a similar excitation spectrum takes place for mercury in the vicinity of the metal-nonmetal transition [13,14]. Such experimental data correspond to the results of molecular-dynamics calculations (see [15,16] and references therein).

\begin{figure}[h]
\centering\includegraphics[width=6cm]{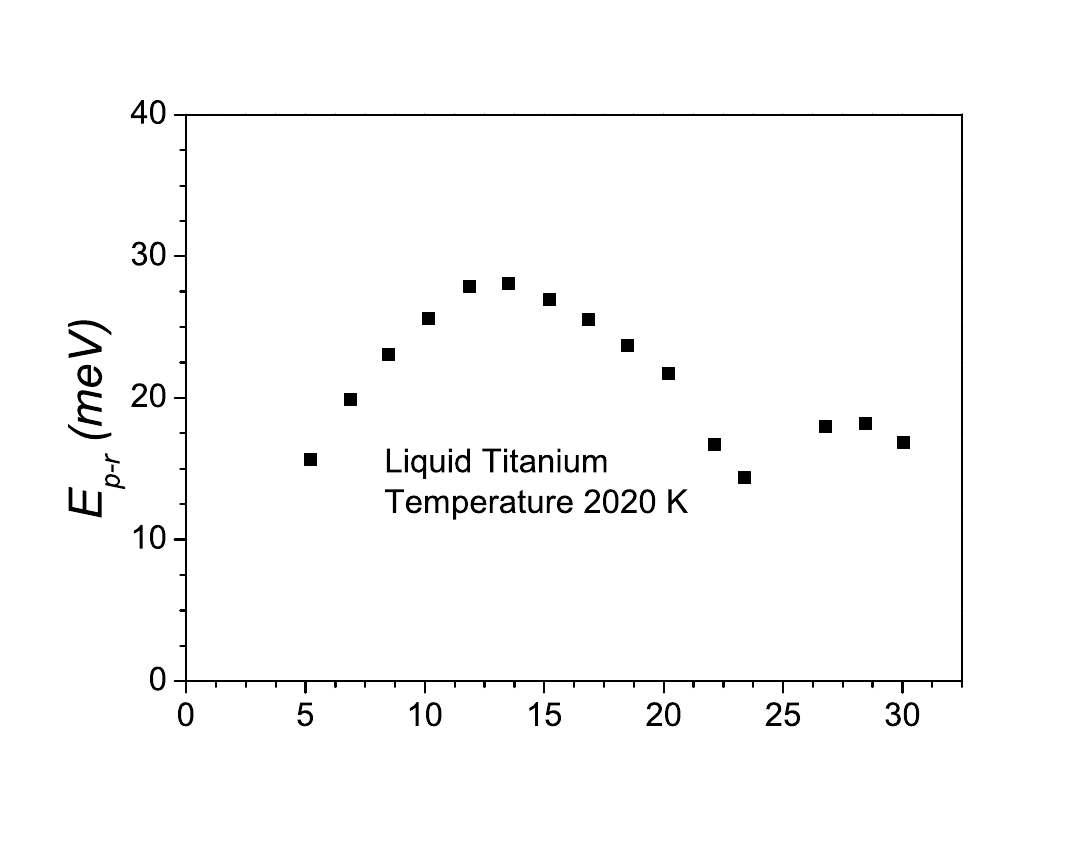}
\caption{{\protect\footnotesize {The curve of maxima of the dynamical structure factors in liquid titanium [12]. }}}\label{Fig.2}
\end{figure}


(B) Nonconductive liquids.
   In [5], the phonon-roton spectrum in normal liquid helium was experimentally observed, and the temperature dependence of the roton minimum $\Delta $ (Fig. 3) was determined; in the experimental study [17], the analogy of the excitation spectrum in superfluid helium and normal liquid was indicated.
   \begin{figure}[h]
\centering\includegraphics[width=6cm]{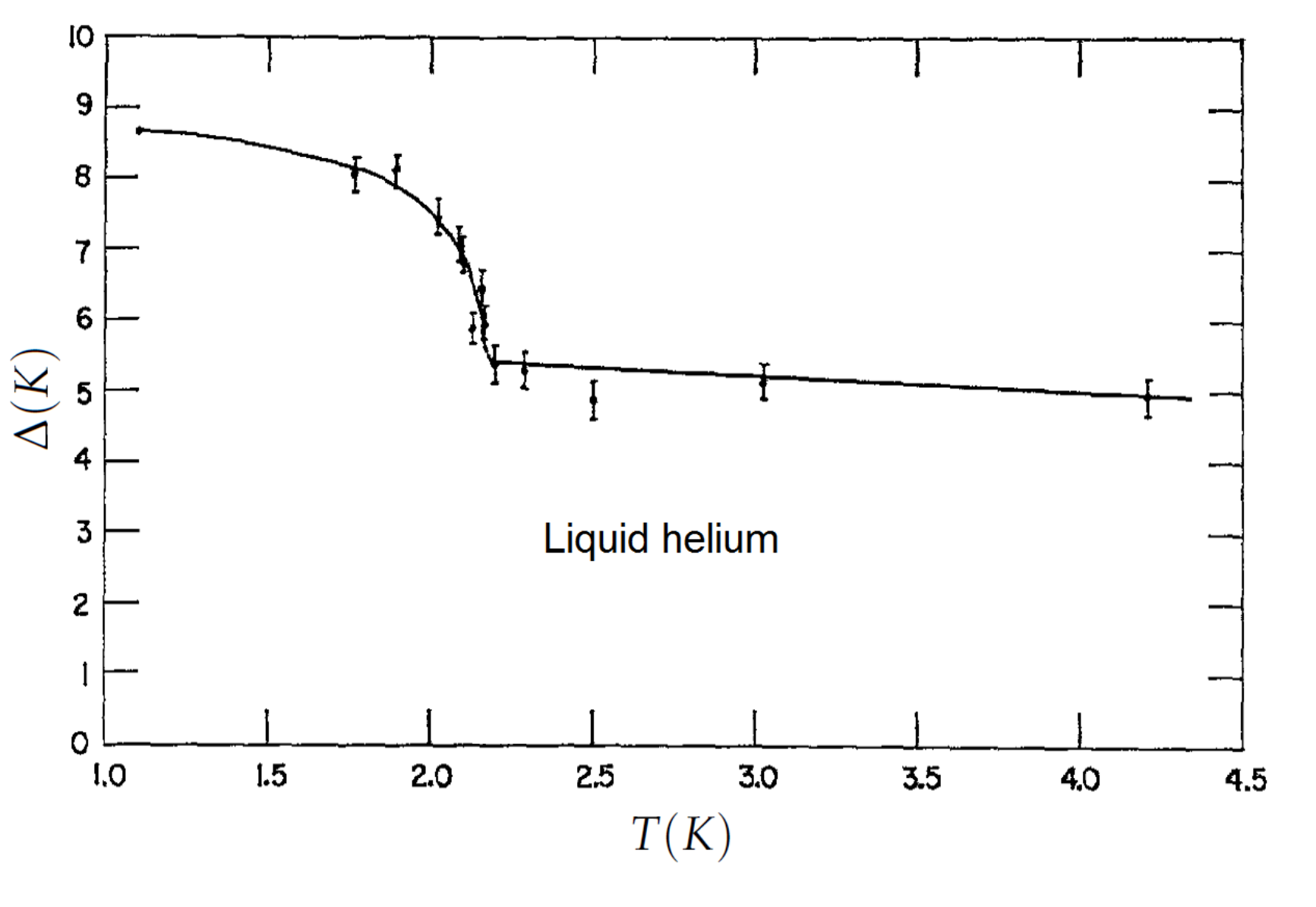}
\caption{{\protect\footnotesize {The experimental temperature dependence of the roton gap in superfluid and normal helium [5]. }}}\label{Fig.3}
\end{figure}


Corresponding experimental data are available also for molecular para-hydrogen [18] (Fig. 4) and neon [19].
 \begin{figure}[h]
\centering\includegraphics[width=6cm]{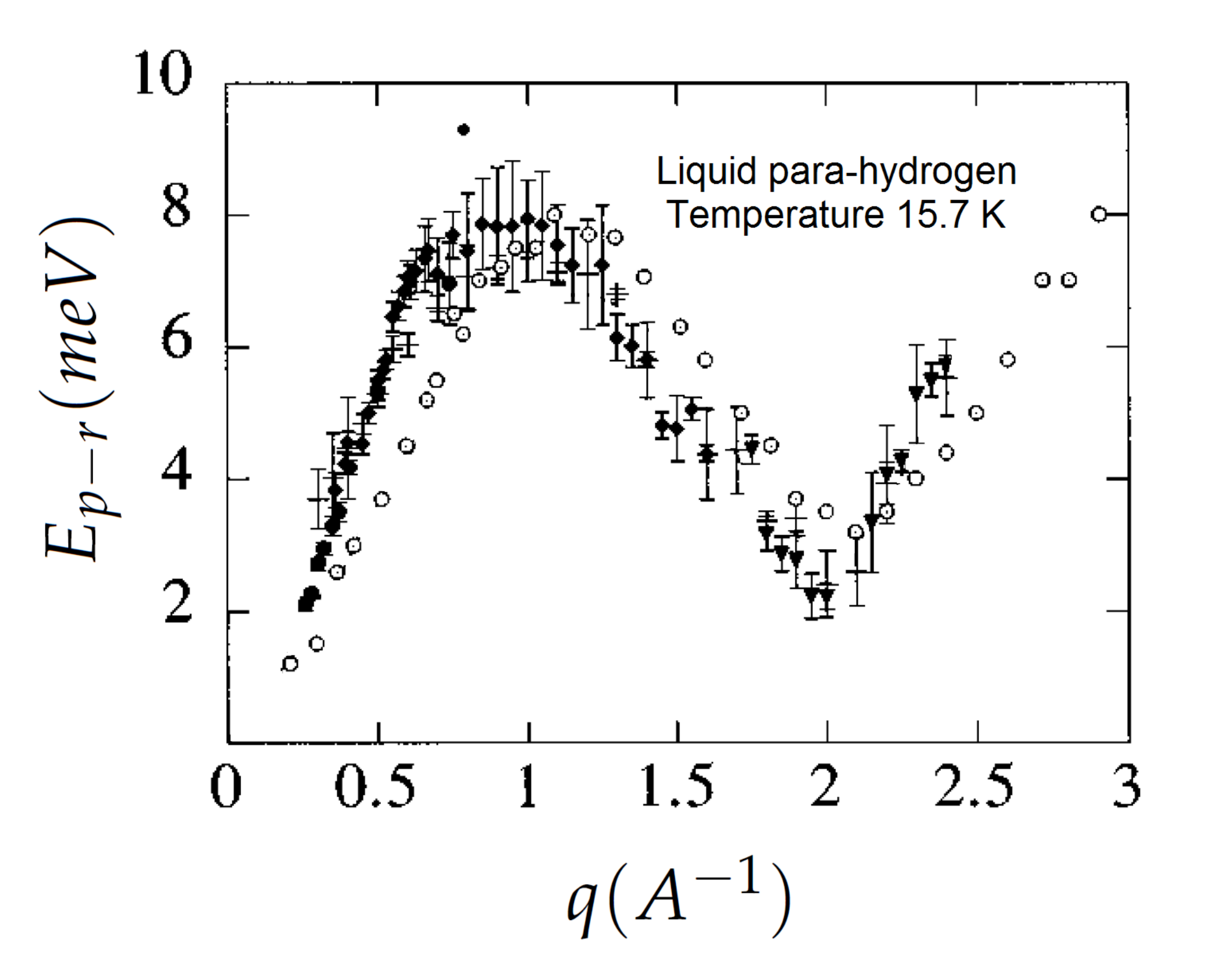}
\caption{{\protect\footnotesize {The experimental INS phonon-roton curve for molecular para-hydrogen at $15.7 K$. The filled symbols (inverted triangles,
diamonds and crosses) depict data measured at different spectrometers and the open circles with a dot show the simulation
results [18].}}}\label{Fig.4}
\end{figure}


For oxygen, similar experimental results were obtained in the supercritical region of thermodynamic parameters [20]. These data are the validation of the assumption (based on an analysis of molecular-dynamic calculations) that the phonon-roton spectrum of collective excitations is characteristic of dense fluids [21]. It should also be noted the possible existence of two branches of phonon-roton-type collective excitations detected in fluid argon [22] in the supercritical range of parameters, which confirms the results of numerical calculations (see [23] and references therein) (Fig. 5).
 \begin{figure}[h]
\centering\includegraphics[width=6cm]{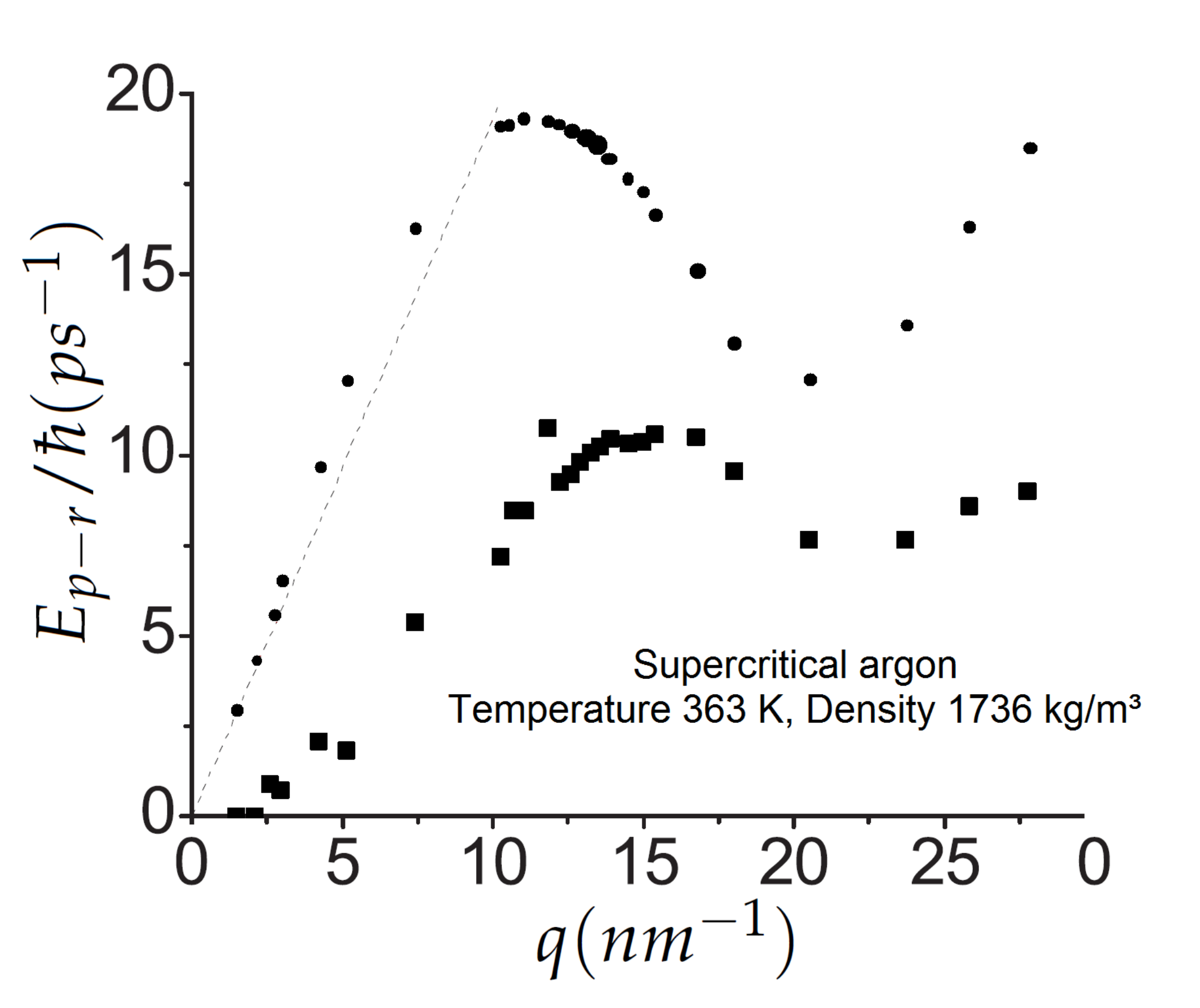}
\caption{{\protect\footnotesize {Dispersion $S(Q, \omega)$ of two types of
collective excitations in supercritical $Ar$ at $T=363 K$ calculated on the basis of Generalized Collective Modes approach [22]. Dispersion of
acoustic excitations (circles) and non-hydrodynamic heat waves (squares) at
density $1249 kg/m^3$. Dashed line corresponds to the hydrodynamic
dispersion. In the long-wavelength limit heat waves do not exist.}}}\label{Fig.5}
\end{figure}


(C) Liquids consisting of several chemical elements.
   In particular, there are experimental data that there exists the phonon-roton spectrum of collective excitations for liquids consisting of very complex molecules of lipid bilayers [24] (Fig. 6) and hydrated $\beta $-lactoglobin [25]. Molecular-dynamic calculations confirm the existence of such a branch of excitations in binary liquids [26, 27].
 \begin{figure}[h]
\centering\includegraphics[width=6cm]{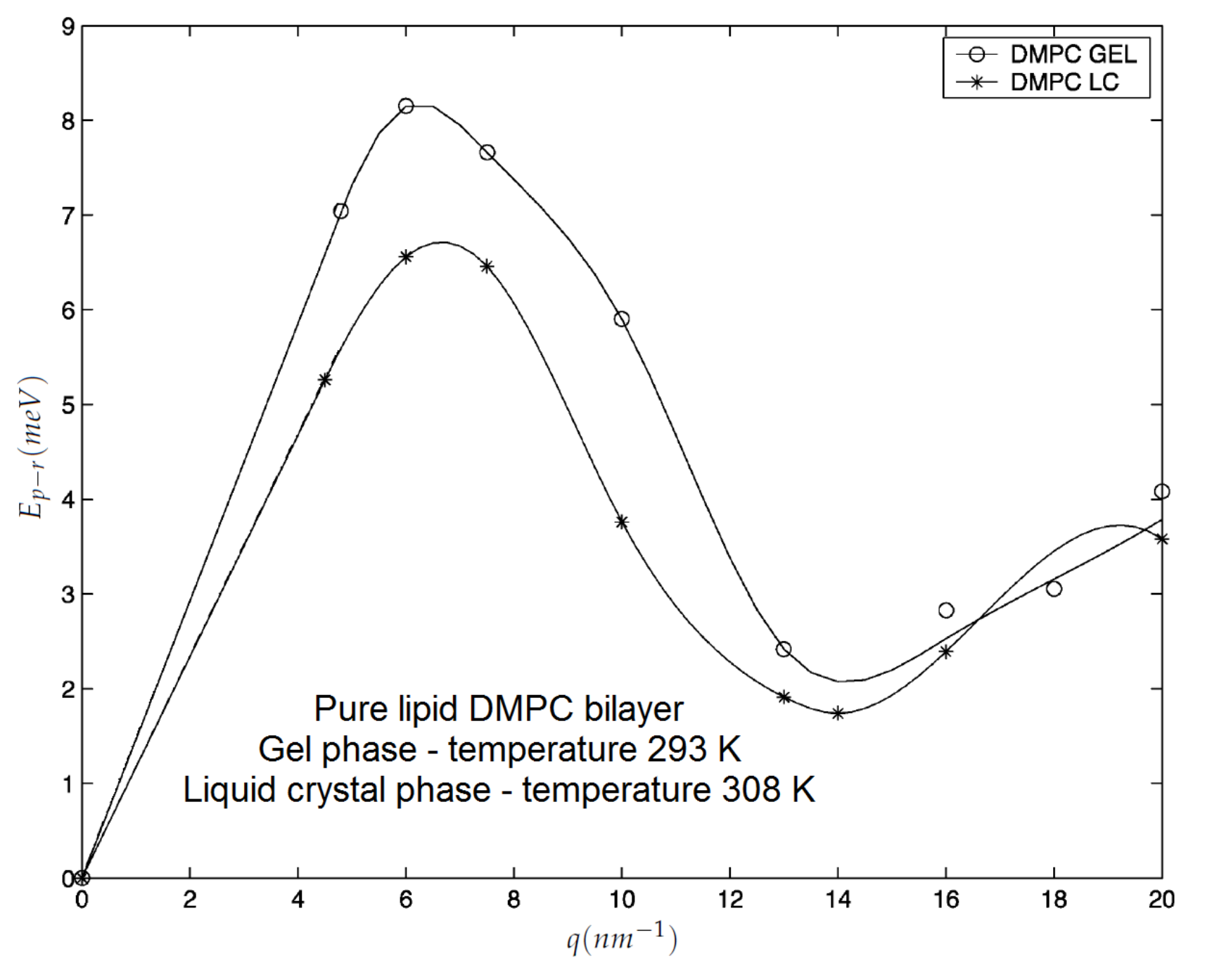}
\caption{{\protect\footnotesize {The experimental (IXS) phonon-roton dispersion curves of dimyristoylphosphatidylcholine (DMPC) bilayer at two temperatures, corresponding to the gel and liquid crystal phases of the bilayer. In the gel
phase ($T=293 K$), the hydrocarbon chains are ordered, so the phonon frequency and the sound speed is higher. In the liquid crystal
phase ($T=308 K$) the chains are disordered, so the phonon frequency and the sound speed are lower. The roton gap at $k=14.0 nm^{-1}$ is also smaller for the liquid crystal case [24].}}}\label{Fig.6}
\end{figure}


(D) Two-dimensional Fermi-liquid.
   The phonon-roton spectrum of collective excitations was obtained based on INS experiments with a two-dimensional layer of liquid $^3$He in the normal state [28].

\section{Universality of phonon-roton excitations: theoretical justification}

Let us now consider the problem of universality of the phonon-roton spectrum of collective excitations in liquids from the viewpoint of theoretical justification. First, we note the general feature in the available experimental data: the position of the roton minimum $p_0$ in the spectrum of collective excitations of liquids qualitatively corresponds to the position of the first maximum in the static structure factor $S(p)$ measured in experiments on elastic neutron and X-ray scattering and directly related to the dynamic structure factor $S(p, E)$ of liquid as
\begin{eqnarray}
n S(p)= \int_{-\infty}^\infty \frac {dE}{2\pi\hbar} S(p,E), \label{A1}
\end{eqnarray}
where $n$ is the density of the number of atoms in the liquid under consideration. Sometimes the wave vector $q=p/\hbar$ and frequency $\omega=E/\hbar$ are used in structure factors as variables. As is known [29], the presence of peaks (maxima) decreasing in magnitude in the static structure factor is caused by the so-called "short-range order"\, of liquids. Therefore, it becomes clear that the existence of the roton minimum in the collective excitation spectrum can be a prominent feature of liquid independently of its physical properties, which corresponds to both experiments and molecular-dynamic calculations mentioned above. This result is in qualitative agreement with the known Feynman formula [30] for the phonon-roton spectrum of collective excitations in superfluid helium
\begin{eqnarray}
E_{p-r}(p)\simeq \frac{p^2}{2mS(p)}. \label{A2}
\end{eqnarray}
We take into account that the existence of corresponding quasiparticles in superfluid helium requires the fulfillment of the following inequalities (see, e.g., [31])
\begin{eqnarray}
E_{p-r}(p)\gg T, \qquad \gamma(p)=\frac {\Gamma(p)}{E_{p-r}(p)} \ll 1.\label{A3}
\end{eqnarray}
Here $T$ is the liquid temperature in energy units, $\gamma(p)$ is the characteristic of phonon-roton excitation damping, which shows the degree of determinacy of the maximum of the dynamic structure factor. The quantity $\hbar/\Gamma(p)$ defines the "lifetime"\, of corresponding quasiparticles, and the function $\Gamma(p)$ is an imaginary part of the pole $\varepsilon(p)$ of the "density-density"\, response function, $\chi(p, E)$ in the lower half-plane of complex $E$,
\begin{eqnarray}
\chi\left(p, E\rightarrow \varepsilon(p)\right)= \rightarrow\infty, \qquad \varepsilon(p)=E(p)-i\Gamma(p).\label{A4}
\end{eqnarray}
The response function $\chi(p, E)$ in the vicinity of the pole $\varepsilon(p)$ (4) can be written as [31]
\begin{eqnarray}
\chi(p,E)=\frac{Z(p)}{E-E_{p-r}(p)+i\Gamma(p)}, \label{A5}
\end{eqnarray}
where $Z(p)$ is a certain unknown function of momentum. In this case, the response function $\chi(p, E)$ is directly related to the dynamic structure factor $S(p, E)$ as
\begin{eqnarray}
S(p,E)=-\frac{2\hbar}{1-\exp(-E/T)} Im \chi(p,E), \qquad S(p,-E)=\exp (-E/T) S(p,E). \label{A6}
\end{eqnarray}
Furthermore, being an analytical function in the upper half-plane of complex $E$, the response function $\chi(p, E)$ satisfies the Kramers-Kronig relations whose direct consequence is the sum rule [31]
\begin{eqnarray}
 \int_{-\infty}^\infty \frac {dE }{2\pi\hbar}E S(p,E)=\frac{n p^2}{2m}. \label{A7}
\end{eqnarray}
As the second condition of (3) is satisfied, the imaginary part of the response function $Im\chi(p, E)$ in the vicinity of the pole $\varepsilon(p)$ can be written in terms of the Dirac $\delta$-function,
\begin{eqnarray}
Im \chi(p,E)= \frac {\Gamma(p)Z(p)}{(E-E_{p-r}(p))^2+\Gamma^2(p)}\approx -\pi Z(p)\delta (E-E_{p-r}(p)) \label{A8}
\end{eqnarray}
   Let the representation (8) be valid for any momentum.
Then, taking into account (5) and substituting (8) into (1) and (7), we come to the Feynman formula (2), if the liquid under consideration contains only one type of quasiparticles (5).  It means, we can apply the Feynman formula to a qualitative analysis of the collective excitation spectrum in normal liquids. In the initial derivation of the formula (2) Feynman used the variational procedure for a Bose system [30] and it can be shown that the spectrum (2) for HeII is always located above the explicit spectrum. This relation is qualitatively  more universal, as follows from the above generalized derivation of the spectrum (3). Generally, for various liquids this spectrum may be placed above as well as below of the real spectrum of excitations.

Therefore, if there are well defined collective excitations in the range of momenta (wave vectors) corresponding to the position of the first maximum of the static structure factor of liquid, the excitation spectrum in this range of momenta is characterized by a roton minimum.
We note that first condition of (3) is not satisfied in the region of extremely small momenta, which corresponds to ordinary sound oscillations. However, there is no doubt that well defined sound collective excitations exist in condensed matter. This means that the Feynman formula (3) at finite temperatures is not valid in the region of extremely small momenta [32].

As a result, we arrive at the conclusion that, at least in the region of not very high temperatures, the phonon-roton spectrum of excitations (in fact, the maxima existence not only for a small $p$, where the sound excitations exist, but also for relatively large $p$, where there is roton maximum in $S(p,E)$) is a universal property of liquids due to the existence of the short-range order, rather than it is a feature of the superfluid state.

\section{New excitations in HeII }

Universality of the phonon-roton excitations opens a new opportunity for description of the superfluid state. First of all, the Landau superfluidity criterion [1] is in fact the superfluidity disappearance criterion of the breakdown of superfluid state, as it was formulated in [33]). Otherwise, taking into account the existence of phonons and rotons in most liquids, we would arrive at the conclusion about the existence of superfluidity in them with the corresponding critical velocity. In other words, the superfluidity criterion is based on the consideration of already existing superfluid flow. Furthermore, universality of the phonon-roton spectrum of collective excitations leads to the statement that rotons are not directly related to the superfluidity phenomenon in ÍåII. This statement is supported, in particular, by the known fact that the critical velocity of superfluidity disappearance in ÍåII, calculated by use the roton minimum is higher than that observed in experiments by orders of magnitude (see [33] for more details). We have stress, however, that according to the existing view the critical velocity determines by the vortexes in superfluid HeII. Existence of vortexes above $T>T_\lambda$ is not enough analyzed and their structure there should be considered in detail separately.

Here, we assume that there are specific excitations of the other type in HeII, which are characteristic exactly of the superfluid state. They can exist only in superfluid helium in addition to phonon-roton excitations and should have the temperature dependence leading to their disappearance at the transition temperature $T_\lambda$, and should describe the measured critical velocities. This hypothesis ensues from the Landau idea about specific role of excitations in superfluid state, which are also crucial for the description of thermodynamic properties. This hypothesis also corresponds to the similar statement for superconductive state where below the transition temperature the specific elementary excitations (Cooper's pairs) appear.
Quantized vortices (see [34] and references therein) can be considered as possible candidates for the role of such excitations.

For this reason, we pay attention that Landau initially considered two types of collective excitations in superfluid helium, i.e., acoustic (characteristic of any liquid) and vortical (rotons, characteristic of superfluid helium) ones [1], which were then combined into a single branch of longitudinal phonon-roton excitations [2]. Feynman implemented the primary Landau's idea about the existence of vortices as those inherent to the superfluid state of quantized excitations [35].
\begin{figure}[h]
\centering\includegraphics[width=6cm]{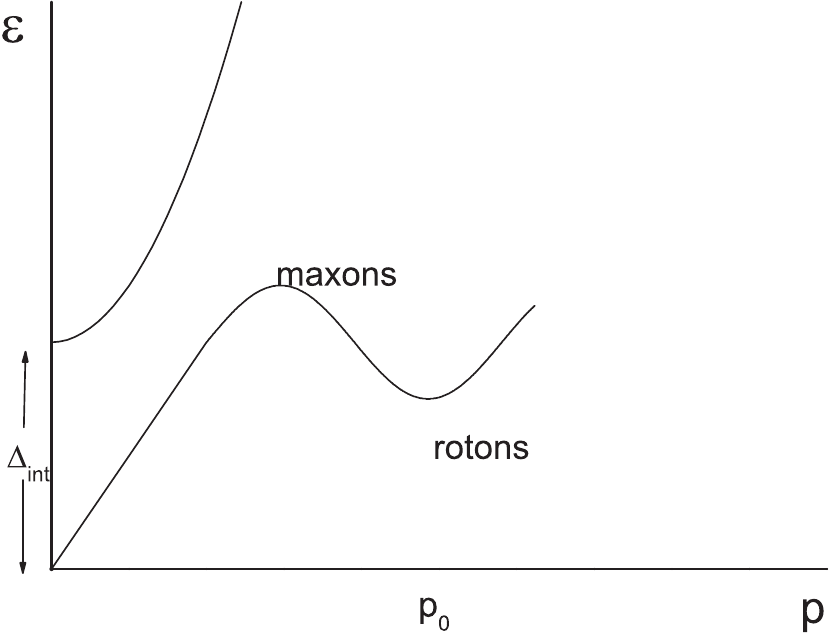}
\caption{{\protect\footnotesize {Phonon--roton spectrum and the expected spectrum of elementary excitations with the gap $\Delta_{int}\equiv \Delta^{(h)}(T)$ (helons [33]) depending on the interaction and temperature at $p = 0$. At the transition temperature, the gap vanishes, $\Delta_{int}(T=T_\lambda)=0$.}}}\label{Fig.7}
\end{figure}

The assumed energy spectrum of excitations, introduced in [33] and called the helons, can be presented as $E(p) = \Delta^{(h)} + p^2/2 m^{(h)}$, where $\Delta^{(h)}(T)$ is the helon gap and $m^{(h)}$ is the corresponding effective mass (see Fig. 7). This spectrum is similar to the roton spectrum first proposed by Landau [1]. The fundamental difference of such quasiparticles from rotons introduced by Landau in [1] is that the gap $\Delta^{(h)}(T)$ is a function of temperature, therewith, $\Delta^{(h)}(T=T_\lambda)=0$ [33]. In this case, the critical velocity $V_{cr} = (2\Delta^{(h)}/m^{(h)})^{1/2}$ of superfluid flow breakdown vanishes at the transition temperature of liquid helium to the normal state. The prominent feature of helons is their disappearance as quasiparticles during the transition of liquid helium to the normal state.

Moreover, as it was shown in [37] (see also [32],[33]) the helons appears at $T\leq T_\lambda$ only as the result of interparticle interaction and in this sense crucially distinguish from the roton spectrum with a gap supposed by Landau in his first paper devoted to HeII [1]. The helon gap disappears when interaction is absent. As it was shown in [32],[33] in the framework of the Landau approach to the thermodynamics of HeII, the existence of the helons gives explanation of the heat capacity singularity at $T\rightarrow T_\lambda$.

\section{Conclusions}

As it was shown in [36] the existence of a gap for small momenta in the energetic spectrum of quasiparticles is not in contradiction with the Goldstone's theorem for quantum liquids with the Bose-Einstein condensate. The question arises: why this branch of excitations, different from the phonon-roton one, does not manifest itself in the INS and IXS experiments? This can be explained by the assumption that these quasiparticles are related with poles of the single-particle Green function, rather with the poles of the "density-density" Green function. These poles coincide in the classical degenerate Bose gas theory [38-40] based on anomalous averages, however, the theory of degenerate Bose gas can be constructed without anomalous averages [32,37, 41-44]. In the seminal paper [38] the anomalous averages have been introduced in Bose gas theory to justify the existence of the phonon-roton excitations as the spectrum of the hamiltonian and to satisfy to the Landau criterion of superfluidity. However, later on, it was shown that the observable phonon-roton excitations are the collective excitations of the dynamical structure factor and there is no necessity to demand their existence and coincidence in the hamiltonian or in single-particle Green function. Therefore, the theory avoiding anomalous averages can be considered at least as alternative opportunity for description of the degenerate Bose gas. This kind of theory should only agree with the general conditions for Bose-condensed system [45,46]. In such type theory the poles of the single-particle Green function and dynamical structure factor are different as for normal systems. All experimental data can be explained on the basis of such type theory if the additional arguments related with fitting of the experimental results to the special theoretical models do not used. Let us accept the most common idea that superfluidity is directly connected with the Bose-Einstein condensation. In this case the helons, which are specific namely for the superfluid state, should be described by the Green function which is most sensitive to the quantum statistics effects. The one-particle Green function, which completely determines the one-particle distribution function over momenta [31], satisfies to this requirement and completely describes the energy of the superfluid state. The anomaly in the momentum distribution function is related with the mathematical description of the Bose-Einstein condensate.

\section*{Acknowledgment}

This study was supported by joint grant of Russian Foundation for Basic Research and Ukraine National Academy of Science, projects No. 12-08-00600-a and No. 12-02-90433-Ukr-a.
S.A. Trigger is thankful to the Netherlands Organization for Scientific Research (NWO) for support of his research by the individual grant in the years 2012-2013.
The authors are thankful to G.J.F. van Heijst, A.M. Ignatov, A.G. Khrapak, A.A. Roukhadze, P.P.J.M. Schram and A.G. Zagorodny for the fruitful discussions and N.M. Blagoveshchenskii for the useful remarks.\\

\end{document}